\documentclass[structabstract]{aa}  
\usepackage{graphicx}
\usepackage{color}
\usepackage{ulem}
\usepackage{booktabs}
\usepackage{natbib}
\bibpunct{(}{)}{;}{a}{}{,} 
\usepackage{subfigure}
\usepackage{txfonts}
\begin{document}
   \title{A homogeneous spectroscopic analysis of host stars of transiting planets\thanks{Based on observations made with the Italian Telescopio Nazionale Galileo (TNG) operated on the island of La Palma by the Fundaci\'on Galileo Galilei of the INAF (Istituto Nazionale di Astrofisica) at the Spanish Observatorio del Roque de los Muchachos of the Instituto de Astrofisica de Canarias}\fnmsep\thanks{Based in part on observations made at Observatoire de Haute Provence (CNRS), France}\fnmsep\thanks{Based on observations made with ESO Telescopes at the La Silla Paranal Observatory under programme ID 080.C-0661.}}
   \author{M. Ammler-von Eiff
          \inst{1,2}
          \and
          N.C. Santos\inst{1}
          \and
          S. G. Sousa \inst{1,3}
          \and
          J. Fernandes \inst{4,5}
          \and
          T. Guillot \inst{6}
          \and
          G. Israelian \inst{7}
          \and
          M. Mayor \inst{8}
          \and
          C. Melo \inst{9}
          }

   \institute{
             Centro de Astrof\'{\i}sica da Universidade do Porto, Rua das Estrelas, 4150-762 Porto, Portugal, \email{matthias@astro.up.pt}         
             \and
Centro de Astronomia e Astrof\'isica da Universidade de Lisboa, Observat\'orio Astron\'omico de Lisboa, Tapada da Ajuda, 1349-018 Lisboa, Portugal
             \and Departamento de Matem\'atica Aplicada, Faculdade de Ci\^encias da Universidade do Porto, Portugal
             \and Centro de F\'{\i}sica Computacional, Universidade de Coimbra, Coimbra, Portugal
             \and Observat\'orio Astron\'omico e Departamento de Matem\'atica, Universidade de Coimbra, Coimbra, Portugal
             \and Observatoire de la C\^ote d'Azur, Laboratoire Cassiop\'ee, CNRS UMR 6202, BP 4229, 06304 Nice Cedex 4, France
             \and Instituto de Astrof\'isica de Canarias, 38200 La Laguna, Tenerife, Spain
             \and Observatoire de Gen\`{e}ve, Universit\'e de Gen\`{e}ve, 51 Ch.des Mailletes, 1290 Sauverny, Switzerland
             \and European Southern Observatory, Casilla 19001, Santiago 19, Chile
             }

   \date{Received ...; accepted ...}

 
  \abstract
   {The analysis of transiting extra-solar planets provides an enormous amount of information about the formation and evolution of planetary systems. A precise knowledge of the host stars is necessary to derive the planetary properties accurately. The properties of the host stars, especially their chemical composition, are also of interest in their own right.}
   {Information about planet formation is inferred by, among others, correlations between different parameters such as the orbital period and the metallicity of the host stars. The stellar properties studied should be derived as homogeneously as possible. The present work provides new, uniformly derived parameters for 13 host stars of transiting planets.}
   {Effective temperature, surface gravity, microturbulence parameter, and iron abundance were derived from spectra of both high signal-to-noise ratio and high resolution by assuming iron excitation and ionization equilibria.}
   {For some stars, the new parameters differ from previous determinations, which is indicative of changes in the planetary radii. A systematic offset in the abundance scale with respect to previous assessments is found for the TrES and HAT objects. Our abundance measurements are remarkably robust in terms of the uncertainties in surface gravities. The iron abundances measured in the present work are supplemented by all previous determinations using the same analysis technique. The distribution of iron abundance then agrees well with the known metal-rich distribution of planet host stars. To facilitate future studies, the spectroscopic results of the current work are supplemented by the findings for other host stars of transiting planets, for a total dataset of 50 objects.}
   {}

\titlerunning{Spectroscopic analysis of stars with transiting planets}
   \maketitle
%

\section{Introduction}

Almost 60 transiting extra-solar planets have so far been found\footnote{See e.g. table at http://www.exoplanet.eu}. This number is small compared to those detections by the radial-velocity technique alone, but the amount of information provided by transiting systems is significantly higher, for the orbits and the structure of the planets. When deriving the planetary properties, a good knowledge of the parent star is important. A sufficiently accurate spectroscopic derivation of effective temperature, surface gravity, and abundances should be based on spectra of both high resolution and high signal-to-noise ratio. 

The knowledge of the chemical abundance of the stars harbouring transiting planets paves the way to interesting studies. \citet{Santos06b} confirmed that they follow the trend of excess metallicity of stars with planets \citep{Gonzalez01,Santos04,Fischer05,Sozzetti09b}.
Heavy elements seem to be important to the formation of close-in planets, as indicated by comparing the planetary heavy-metal content with the metallicity of the host star \citep{Guillot06}. This work will benefit strongly from more accurate and uniform parameters of host stars of transiting planets.

Stellar parameters and chemical abundances have already been derived for most parent stars of transiting planets, usually in a very uniform way within each transit search program, e.g., as in the case of the HAT objects \citep{Bakos07a,Bakos07b,Kovacs07,Noyes08,Pal08a} and the TrES stars \citep{Sozzetti06,Sozzetti07,Sozzetti09a}. These data have been used to homogeneously reassess and interpret the properties of the entire list of transiting planets \citep{Torres08,Southworth08,Southworth09}. However, it is difficult to take into account the underlying stellar parameters in a uniform way. \citet{Torres08} note that measurements for the same star may not agree, and that systematic errors may be present, which are treated differently. They decided to critically reassess the values and their error bars based on the information given in the literature. 

Systematics may result from, e.g., issues of data reduction, the choice of different analysis tools, line lists, and oscillator strengths. From Galactic studies, it is well known that different spectroscopic studies of the same stars do not necessarily provide consistent stellar parameters and abundances, as is also found in extra-solar planet research, e.g., the case of OGLE-TR-10 \citep{Ammler08}. At the same time, published error bars do not necessarily account for the systematics involved and usually underestimate the true uncertainties. The reader is referred to the discussion in \citet{Torres08} and references given therein, as well as to \citet{Fuhrmann98} and \citet{Luck05}.

The situation is improved by a homogeneous spectroscopic analysis that is as accurate as possible for the entire sample of parent stars. \citet{Santos06b} studied the host stars OGLE-TR-10, 56, 11, 113, and TrES-1. Based on high-resolution spectra with a signal-to-noise ratio of at least $\approx100$, they measured equivalent widths of lines of neutral (Fe\,I) and ionized iron (Fe\,II). Imposing the excitation and ionization equilibrium, they derived effective temperature, surface gravity, and iron abundance [Fe/H].

The goal of this work is two-fold and continues the work of \citet{Santos06b}. Firstly, accurate stellar parameters of the host stars of 13 transiting planets are derived homogeneously and combined with previous measurements for 11 further host stars, measurements that were obtained in the same way. Therefore, this work presents an extensive data set of homogeneously derived stellar parameters and iron abundances of parent stars of transiting planets. Secondly, the metallicity distribution of stars harbouring transiting planets is readdressed using the new homogeneous compilation of iron abundances. 

In addition, all available spectroscopic parameters of known host stars of transiting planets are compiled, providing a possible starting point for future studies of the properties of planets and their host stars, particularly those that involve chemical abundances.

\section{Observations and data reduction}

\begin{table}
\centering
\caption{\label{tab:obs}\textbf{Observations log} - Only the total exposure times are given. The exposure times of the coadded individual spectra are mentioned in the text.}
\begin{tabular}{llrr}
\toprule
target      &instrument&$t_\mathrm{exp}[\mathrm{s}]$&S/N\\
\midrule
CoRoT-Exo-2&UVES  & 2,400              &200\\
HAT-P-1    &SARG  &18,000              &210\\
HAT-P-2    &SARG  & 7,200              &340\\
HAT-P-4    &SOPHIE&11,000              &80\\
HAT-P-6    &SOPHIE&18,000              &180\\
HAT-P-7    &SOPHIE&14,400              &150\\
HD\,17156 &SOPHIE& 3,600              &220\\
HD\,149026&SARG  & 7,200              &420\\   
TrES-2    &SARG  &21,600              &200\\
TrES-3    &SARG  &28,800              &100\\
TrES-4    &SOPHIE&14,400              &90\\
XO-1      &SARG  &25,200              &170\\
XO-2      &SOPHIE&18,000              &110\\
\bottomrule
\end{tabular}
\end{table}

High resolution spectra of six stars were obtained by the SARG spectrograph  at the TNG, La Palma, and of six additional stars with SOPHIE at the OHP, France. In addition, UVES spectra of one more star, CoRoT-Exo-2, were taken from the ESO archive. The data were reduced in different ways, as explained in detail below. Afterwards, radial velocity shifts were corrected using the IRAF\footnote{IRAF is distributed by National Optical Astronomy Observatories, operated by the Association of Universities for Research in Astronomy, Inc., under contract with the National Science Foundation, USA.} task DOPCOR and individual exposures were added using the task SCOMBINE.

\subsection{SARG}

The spectral resolving power of the SARG spectra is $\frac{\lambda}{\Delta\lambda}\approx57,000$. The configuration of SARG comprises the yellow grism and the yellow rejection filter covering the wavelength range $4650$-$7920\,\mathrm{\AA}$. However, there is a gap at $6110$-$6350\,\mathrm{\AA}$, since the mosaic detector consists of two CCD's with 2Kx4K pixels. A read-out binning of 2x1 was chosen. The observations were carried out during July 21-24, 2007. For every star, several spectra each with an exposure time of 1800\,s were taken. Table~\ref{tab:obs} lists the total exposure time for each star. 
The individual SARG frames were reduced with the IRAF tasks APALL, DISPCOR, and ECIDENTIFY. 
All spectra are of high signal-to-noise ratio of at least 100 as measured in small spectral windows between 6420 and 6430\,{\AA}.

\subsection{SOPHIE}

The spectral resolving power of SOPHIE \citep{Perruchot08} is somewhat higher using the high-resolution mode ($\frac{\lambda}{\Delta\lambda}\approx75,000$) and covers a wavelength range of $3870$-$6940\,\mathrm{\AA}$. The observations were carried out during Oct 26-28, 2008 and were affected by variable conditions. It was still possible to subtract the background by using the second fiber that was targeted at a sky position. Typical exposure times of 1800\,s and 3600\,s were used for individual frames. Table~\ref{tab:obs} again indicates the total exposure time for each star. The SOPHIE spectra were reduced with the on-site reduction pipeline using optimal extraction. The signal-to-noise ratios provided in the table were calculated by adding in quadrature the values given by the pipeline reduction for each individual frame. The signal-to-noise ratio achieved was generally lower than in the case of the SARG spectra but still of the order of 100 and higher. 

\subsection{UVES}

Two spectra each with exposure times of 1200\,s each were acquired for CoRoT-Exo-2 with UVES \citep{Dekker00} on Oct 13, 2007. The cross-disperser \#3 with a central wavelength of 580.0\,nm was used providing a wavelength coverage of $4200$-$6800\,\mathrm{\AA}$. A slit width of $0{\farcs}7$ and a binning of 1x1 was chosen implying a resolving power of about $60,000$. These observations form part of the programme ID 080.C-0661 and were retrieved from the ESO archive. The raw data were reduced with a locally installed version (2.9.7) of the reduction pipeline in combination with the MIDAS version 07SEPpl1.1. The red arm spectra of both the upper and lower CCD were optimally extracted, preserving the wavelength range. A section of $70\,\mathrm{\AA}$ is missing at the central wavelength because of the gap between the two detectors. The signal-to-noise ratio given in Table~\ref{tab:obs} was estimated in the same way as for the SARG spectra.

\section{Derivation of stellar parameters and metallicity}

The work of \citet{Santos06b} is extended consistently by analysing the spectra in the same way as described in \citet{Santos04}. In brief, the analysis begins with the selection of weak iron lines from a list of 39 Fe\,I and 16 Fe\,II lines. While SOPHIE and UVES detect almost all lines in this list, the detector gap of the SARG spectrograph happens to correspond to a wavelength region that is rich in useful Fe lines. Therefore, the analysis of the SARG spectra has to rely typically on only 20 Fe\,I lines and 9 Fe\,II lines. 

Equivalent widths were measured with the IRAF SPLOT task contained in the ECHELLE package. Effective temperature, surface gravity, and iron abundance [Fe/H] were derived for local thermal equilibrium by assuming iron excitation and ionization equilibrium. The correct value of the microturbulence parameter was chosen by ensuring that [Fe/H] does not vary with reduced equivalent width. The procedure is based on the use of the 2002 version of MOOG\footnote{http://verdi.as.utexas.edu/moog.html} \citep{Sneden73} and ATLAS plane-parallel model atmospheres \citep{Kurucz93}.

Error bars were derived in the same way as described in \citet{Santos04} and \citet{Santos06b}, who follow the method of \citet{Gonzalez96} and \citet{Gonzalez98}. The error bars are purely statistical, consisting of the dispersion in the iron abundances as measured from each iron line, as well as the uncertainties in the slopes of the correlations of iron abundance with reduced equivalent width, excitation, and ionization potential.

The derived stellar parameters are presented in the upper part of Table~\ref{tab:par}, which in addition includes data for stars harbouring transiting planets analysed by the same method. In the lower part of the table, all other host stars of transiting planets\footnote{According to http:www.exoplanet.eu.} with available abundance information are compiled. Preference is given to spectroscopically measured effective temperatures and surface gravities, since these represent the basis of spectroscopic abundance measurements. In some cases however, the analysis was not based exclusively on spectroscopic data. The most reliable spectroscopic metallicity measurements are compiled for future work, e.g., studies of correlations of planetary properties with the properties of the parent stars.

\begin{table*}
   \caption{\label{tab:par}\textbf{Stellar parameters and iron abundance of host stars of transiting planets.} -- The upper part of the table lists the parameters homogeneously derived using the method of \citet{Santos04}. The lower part contains data for all host stars of transiting planets not studied in this work but with stellar parameters and iron abundance or metallicity published elsewhere.}
      \begin{tabular}{lccccc@{,}cc@{,}cl}
      \toprule
      Star&$T_\mathrm{eff}$\,[K]&${\log}g\,$(cgs)&$\xi_\mathrm{t}$\,[km$\mathrm{s}^{-1}]$&\multicolumn{1}{c}{[Fe/H]}&\multicolumn{2}{c}{N(Fe\,I, Fe\,II)}&\multicolumn{2}{c}{$\sigma(\mathrm{Fe\,I, Fe\,II})$}&Source\\
      \midrule
      \object{CoRoT-Exo-2}&$5608\pm37$&$4.71\pm0.20$&$1.49\pm0.06$&$0.07\pm0.04 $&26&9&0.03&0.09&this work\\
      \object{HAT-P-1}     &$6076\pm27$&$4.47\pm0.07$&$1.17\pm0.05$&$ 0.21\pm0.03$&21&10&0.02&0.03&this work\\
      \object{HAT-P-2}    &$6951\pm181$&$4.45\pm0.08$&$1.61\pm0.33$&$0.49\pm0.12$&14&8&0.06&0.03&this work\\
      \object{HAT-P-4}    &$6054\pm 60$&$4.17\pm0.28$&$1.59\pm0.09$&$0.35\pm0.08$&30&9&0.06&0.12&this work\\
      \object{HAT-P-6}    &$6855\pm111$&$4.69\pm0.20$&$2.85\pm1.15$&$-0.08\pm0.11$&21&9&0.07&0.07&this work\\
      \object{HAT-P-7}    &$6525\pm 61$ &$4.09\pm0.08$&$1.78\pm0.14$&$0.31\pm0.07$&33&10&0.06&0.03&this work\\
      \object{HD\,17156} &$6084\pm 29$&$4.33\pm0.05$&$1.47\pm0.05$&$0.23\pm0.04$&30&10&0.03&0.02&this work\\
      HD\,80606 &$5574\pm72$&$4.46\pm0.20$&$1.14\pm0.09$&$0.32\pm0.09$&38&5&0.07&0.08&\citet{Santos04}\\
      \object{HD\,149026}&$6162\pm41$&$4.37\pm0.10$ &$1.41\pm0.07$&$0.36\pm0.05$&23&9&0.04&0.04&this work\\
      HD\,189733&$5050\pm50$&$4.53\pm0.14$&$0.95\pm0.07$&$-0.03\pm0.04$&--&--&--&--&\citet{Bouchy05b}\\
      HD\,209458&$6117\pm26$&$4.48\pm0.08$&$1.40\pm0.06$&$0.02\pm0.03$&--&--&--&--&\citet{Santos04}\\
      OGLE-TR-10&$6075\pm86$&$4.54\pm0.15$&$1.45\pm0.14$&$0.28\pm0.10$&33&11&0.08&0.06&\citet{Santos06b}\\
      OGLE-TR-56&$6119\pm62$&$4.21\pm0.19$&$1.48\pm0.11$&$0.25\pm0.08$&31&9&0.06&0.08&\citet{Santos06b}\\
      OGLE-TR-111&$5044\pm83$&$4.51\pm0.36$&$1.14\pm0.10$&$0.19\pm0.07$&31&7&0.07&0.18&\citet{Santos06b}\\
      OGLE-TR-113&$4804\pm106$&$4.52\pm0.26$&$0.90\pm0.18$&$0.15\pm0.10$&30&5&0.10&0.09&\citet{Santos06b}\\
      OGLE-TR-132&$6411\pm179$&$4.86\pm0.14$&$1.46\pm0.36$&$0.43\pm0.18$&--&--&--&--&\citet{Bouchy04}\\
      OGLE-TR-182&$5924\pm64$&$4.47\pm0.18$&--&$0.37\pm0.08$&--&--&--&--&\citet{Pont08}\\
      OGLE-TR-211&$6325\pm91$&$4.22\pm0.17$&$1.63\pm0.21$&$0.11\pm0.10$&--&--&--&--&\citet{Udalski08}\\
      TrES-1     &$5226\pm38$&$4.40\pm0.10$&$0.90\pm0.05$&$ 0.06\pm0.05$&36&7&0.04&0.05&\citet{Santos06b}\\
      \object{TrES-2}     &$5795\pm73$&$4.30\pm0.13$&$0.79\pm0.12$&$ 0.06\pm0.08$&20&9&0.06&0.05&this work\\
      \object{TrES-3}     &$5502\pm157$&$4.44\pm0.22$&$1.00\pm0.30$&$-0.10\pm0.19$&20&8&0.14&0.07&this work\\
      \object{TrES-4}     &$6293\pm 96$&$4.20\pm0.27$&$2.01\pm0.17$&$0.34\pm0.10$&27&10&0.07&0.11&this work\\
      \object{XO-1}       &$5754\pm42$ &$4.61\pm0.05$&$1.07\pm0.09$&$-0.01\pm0.05$&21&9&0.04&0.02&this work\\
      \object{XO-2}       &$5350\pm72$&$4.14\pm0.22$&$1.10\pm0.08$&$0.42\pm0.07$&22&9&0.05&0.10&this work\\
      \midrule
      CoRoT-Exo-1&$5950\pm150$&$4.25\pm0.30$&&$-0.3\pm0.25^{a}$&\multicolumn{5}{l}{\citet{Barge08}}\\
      CoRoT-Exo-3&$6740\pm140$&$4.22\pm0.07$&&$0.02\pm0.06^{f}$&\multicolumn{5}{l}{\citet{Deleuil08}}\\
      CoRoT-Exo-4&$6190\pm60 $&$4.41\pm0.05$&$0.94\pm0.07$&$0.05\pm0.07^{a}$&\multicolumn{5}{l}{\citet{Moutou08}}\\
      GJ\,436    &$3350\pm300$&$5.0$&&$-0.02\pm0.20$$^e$&\multicolumn{5}{l}{\citet{Bonfils05,Maness07}}\\
      HAT-P-3     &$5185\pm46$ &$4.61\pm0.05$&&$0.27\pm0.04$&\multicolumn{5}{l}{\citet{Torres07}}\\
      HAT-P-5     &$5960\pm100$&$4.368\pm0.028$&&$0.24\pm0.15$&\multicolumn{5}{l}{\citet{Bakos07c}}\\
      HAT-P-8     &$6200\pm80$&$4.15\pm0.03$&&$0.01\pm0.08$&\multicolumn{5}{l}{\citet{Latham08}}\\
      HAT-P-9     &$6350\pm150$&$4.29^{+0.03}_{-0.04}$&&$0.12\pm0.20$&\multicolumn{3}{l}{\citet{Shporer09}}\\
      HAT-P-11    &$4780\pm50$&$4.59\pm0.03$&&$0.31\pm0.05$&\multicolumn{5}{l}{\citet{Bakos09b}}\\
      OGLE2-TR-L9&$6933\pm58$&$4.47\pm0.13$&&$-0.05\pm0.20$&\multicolumn{5}{l}{\citet{Snellen09}}\\
      WASP-1     &$6110\pm45$ &$4.28\pm0.15$&&$0.26\pm0.03^{b}$     &\multicolumn{3}{l}{\citet{Stempels07}}\\
      WASP-2     &$5200\pm200$&$4.3\pm0.3$&&$0.1\pm0.2$&\multicolumn{5}{l}{\citet{Cameron07}}\\
      WASP-3     &$6400\pm100$&$4.25\pm0.05$&&$0.00\pm0.20^{a}$&\multicolumn{5}{l}{\citet{Pollacco08}}\\
      WASP-4     &$5500\pm150$&$4.3\pm0.2$&&$0.0\pm0.2^{a}$&\multicolumn{5}{l}{\citet{Wilson08}}\\
      WASP-5     &$5700\pm150$&$4.3\pm0.2$&&$0.0\pm0.2^{a}$&\multicolumn{5}{l}{\citet{Anderson08}}\\
      WASP-6     &$5450\pm100$&$4.6\pm0.2$&$1.0\pm0.2$&$-0.20\pm0.09$&\multicolumn{5}{l}{\citet{Gillon09}}\\
      WASP-7     &$6400\pm100$&$4.3\pm0.2$&$1.5\pm0.2$&$0.0\pm0.1$&\multicolumn{5}{l}{\citet{Hellier09}}\\
      WASP-10    &$4675\pm100$&$4.40\pm0.20$&&$0.03\pm0.20^{a}$&\multicolumn{5}{l}{\citet{Christian09}}\\
      WASP-11/HAT-P-10    &$4980\pm60$&$4.54\pm0.03$&&$0.13\pm0.08$&\multicolumn{5}{l}{\citet{Bakos09a}}\\
      WASP-12    &$6300^{+200}_{-100}$&$4.38\pm0.10$&&${0.30^{+0.05}_{-0.15}}^{a}$&\multicolumn{5}{l}{\citet{Hebb09}}\\
      {\bf WASP-13}    &$5826\pm100$&$4.04\pm0.2 $&&$0.0\pm0.2^{a}$&\multicolumn{5}{l}{\citet{Skillen09}}\\
      WASP-14    &$6475\pm100$&$4.07\pm0.20$&&$0.0\pm0.2^{a}$&\multicolumn{5}{l}{\citet{Joshi09}}\\
      WASP-15    &$6300\pm100$&$4.35\pm0.15$&$1.4\pm0.1$&$-0.17\pm0.11$&\multicolumn{5}{l}{\citet{West09}}\\
      XO-3       &$6429\pm100$&$4.244\pm0.041$&&$-0.177\pm0.080^{c}$&\multicolumn{5}{l}{\citet{Winn08}, updating \citet{Johns-Krull08}}\\
      XO-4       &$6397\pm70$ &$4.18\pm0.07$&&$-0.04\pm0.03^{d}$&\multicolumn{5}{l}{\citet{McCullough08}}\\
      XO-5       &$5510\pm44$ &$4.52\pm0.06$&&$0.25\pm0.03$&\multicolumn{5}{l}{\citet{Burke08}}\\
      {\bf XO-5}       &$5370\pm70$ &$4.31\pm0.03$&&$0.05\pm0.06$&\multicolumn{5}{l}{\citet{Pal09}}\\
      \bottomrule
   \end{tabular}\\
      $^{a}$ The cited value gives overall metallicity [M/H] rather than [Fe/H].
      $^{b}$ They provide a value for overall metallicity of $0.23\pm0.08$.
      $^{c}$ They provide a value for overall metallicity of $-0.204\pm0.023$.
      $^{d}$ They provide a value for overall metallicity of $-0.02\pm0.05$.
      $^{e}$ The average of the values given by \citet{Bonfils05} and \citet{Maness07} is adopted here.
      $^{f}$ The abundance derived from Fe\,II lines is given. \citet{Deleuil08} infer an overall metallicity of [M/H]$=-0.02\pm0.06$.
\end{table*}

\begin{table*}
   \caption{\label{tab:others}\textbf{Stellar parameters and metallicities previously derived by spectroscopy.} Only the stars analysed in the present work are listed. The last column repeats the values from Table~\ref{tab:par} to facilitate the comparison.}
      \begin{tabular}{lrccclc}
      \toprule
      Star&$T_\mathrm{eff}$\,[K]&${\log}g\,$(cgs)&$\xi_\mathrm{t}$\,[km$\mathrm{s}^{-1}]$&[Fe/H]&Source&[Fe/H] (this work)\\
      \midrule
      CoRoT-Exo-2&$5625\pm120$&$4.3\pm0.2  $ &--&$0.0\pm0.1$ &\citet{Bouchy08}&$+0.07\pm0.04$\\
      HAT-P-1    &$5795\pm45$&$4.45\pm0.06$ &--&$+0.13\pm0.02$ &\citet{Bakos07a}&$+0.21\pm0.03$\\
      HAT-P-2    &$6290\pm110$&$4.22\pm0.14$&--&$+0.12\pm0.08$ &\citet{Bakos07b}&$+0.49\pm0.12$\\
                &    --     &       --     &--&$+0.11\pm0.10$ &\citet{Loeillet08}&\\
      HAT-P-4    &$6032\pm80$&$4.36\pm0.11$&--&$+0.32\pm0.08$&\citet{Kovacs07}&\\
                &$5860\pm80$&$4.14^{+0.01}_{-0.04}$&--&$+0.24\pm0.08$&\citet[using isochrones]{Kovacs07}&$+0.35\pm0.08$\\
      HAT-P-6    &$6353\pm88$&$3.84\pm0.12$&--&$-0.23\pm0.08$&\citet{Noyes08}&\\
                &$6570\pm80$&$4.22\pm0.03$&--&$-0.13\pm0.08$&\citet[using isochrones]{Noyes08}&$-0.08\pm0.11$\\
      HAT-P-7    &$6350\pm80$&${4.06\pm0.10}^a$&--&$+0.26\pm0.08$&\citet{Pal08a}&$+0.31\pm0.07$\\
      HD\,17156 &$6079\pm56$&$4.29\pm0.06$&--&$+0.24\pm0.03$&\citet{Fischer07}&$+0.23\pm0.04$\\
      HD\,149026&$6147\pm50$&$4.26\pm0.07$ &--&$+0.36\pm0.05$ &\citet{Sato05}&$+0.36\pm0.05$\\
      TrES-2    &$5850\pm50$&$4.4\pm0.1$   &$1.00\pm0.05$&$-0.15\pm0.10$&\citet{Sozzetti07}&$+0.06\pm0.08$\\
      TrES-3    &$5650\pm75$&$4.4\pm0.1$   &$0.85\pm0.05$&$-0.19\pm0.08$&\citet{Sozzetti09a}&$-0.10\pm0.19$\\
      TrES-4    &$6200\pm75$&$4.0\pm0.1$   &$1.50\pm0.05$&$+0.14\pm0.09$&\citet{Sozzetti09a}&$+0.34\pm0.10$\\
      XO-1      &$5750\pm13$&$4.53\pm0.065$&--&$+0.015\pm0.04$&\citet{McCullough06}&$-0.01\pm0.05$\\
      XO-2      &$5340\pm32$&$4.48\pm0.05$ &--&$+0.45\pm0.02$&\citet{Burke07}&$+0.42\pm0.07$\\
      \bottomrule
   \end{tabular}\\
$^a$ virtually identical to the value of $4.07^{+0.04}_{-0.08}$ derived using isochrones.
\end{table*}

\section{Comparison with previous measurements}

\citet{Torres08} compiled literature values for host stars of transiting planets, including those in the present work. To refine planetary parameters, \citet{Torres08} revised the stellar data from the literature. In contrast, the present work aims to provide an improved set of stellar parameters based on spectroscopy, then compare with previous determinations. The remarks made by \citet{Torres08} concerning the limits of this comparison have to be taken into account. In this respect, a discussion of the spectroscopic techniques will be important.

The parameters are compared with those determined by other groups (Table~\ref{tab:others}). The comparison is restricted to spectral synthesis studies, although other measurements may have been determined by other methods. For example, in the case of HD\,149026, the semi-empirical temperature determination of \citet{Masana06} of $6183\pm57$\,K agrees well with the spectroscopic measurements. Consistent results are also obtained by \citet{Robinson07} using Lick indices ($T_\mathrm{eff}=6081\,\mathrm{K}$, ${\log}g=4.24$, $\mathrm{[Fe/H]}=0.32$), although they state that these parameters are beyond any of the tested ranges.

\subsection{Only few different techniques}
In terms of the methods used for spectroscopic analysis, the previous studies are very homogeneous, i.e., only a few different techniques were used by all groups. \citet{Bakos07a}, \citet{Bakos07b}, \citet{Burke07}, \citet{Fischer07}, \citet{Kovacs07}, \citet{McCullough06}, \citet{Noyes08}, \citet{Pal08a}, and \citet{Sato05} all of which are referred to \citet{Valenti05}, mostly explicitly state the use of the SME (Spectroscopy Made Easy) software \citep{Valenti96}, which is based on ATLAS model atmospheres. This spectral fitting technique varies the stellar parameters and the abundances of some species to minimize the differences between the synthetic spectra and observed spectrum. \citet{Kovacs07}, \citet{Noyes08}, and \citet{Pal08a} refined surface gravity using model isochrones and repeated the analysis iteratively as required. \citet{Bouchy08} also applied a spectral fitting technique, using MARCS model atmospheres.

The approach of \citet{Sozzetti07,Sozzetti09a}, however, is very similar to the present work imposing the Fe\,I/II ionization and excitation equilibria using the MOOG line formation code and ATLAS model atmospheres. A different technique was used by \citet{Loeillet08}, who analysed the spectra based on a cross-correlation technique following \citet{Santos02}. 

Some groups supplement their measurements with results from other techniques, e.g., using H$\alpha$ and line depth ratios \citep{Bouchy08,Sozzetti07,Sozzetti09a}.

\subsection{Small amount of variety between spectrographs used}
The spectra used by the different groups to derive stellar parameters were taken with different instruments, and had different spectral resolutions and signal-to-noise ratios. 
Large differences because of the use of different instruments are not expected, since all are \'Echelle spectrographs with high resolving powers. Half of the studies mentioned in Table~\ref{tab:others} are based on spectra taken with the Keck HIRES spectrograph, although with slightly different configurations. \citet{Bakos07a}, \citet{Bakos07b}, \citet{Kovacs07}, \citet{Noyes08}, and \citet{Pal08a} obtained Keck HIRES spectra with a resolving power of $\frac{\lambda}{\Delta\lambda}=55,000$, while the HIRES spectra used by \citet{Sozzetti07,Sozzetti09a} have $\frac{\lambda}{\Delta\lambda}=71,000$. \citet{Burke07} and \citet{McCullough06} obtained spectra with $\frac{\lambda}{\Delta\lambda}=60,000$ with the \'Echelle spectrograph at the Harlan J. Smith telescope. The results of \citet{Sato05} are based on SUBARU HDS spectra with $\frac{\lambda}{\Delta\lambda}=55,000$. The spectra analysed by \citet{Loeillet08} were obtained with the SOPHIE spectrograph. 

A comparison of the signal-to-noise ratios is not possible since most groups do not explicitly state the precise values. \citet{Sato05} indicate a typical value of $\approx150/\mathrm{pix}$ at $5500\,\mathrm{\AA}$ for all their stars. The HARPS spectra used by \citet{Bouchy08} have a signal-to-noise ratio of about $80/\mathrm{pix}$. \citet{Sozzetti07} obtained an average of $120/\mathrm{pix}$ for TrES-2, while \citet{Sozzetti09a} achieved $100/\mathrm{pix}$.

The present work is based on spectra of high signal-to-noise ratio that were observed, reduced, and analysed in the same way. Therefore, the new results allow us to check the consistency of previously obtained results.

\subsection{Systematics and possible explanations}

\begin{figure*}
\subfigure{\includegraphics[width=88mm]{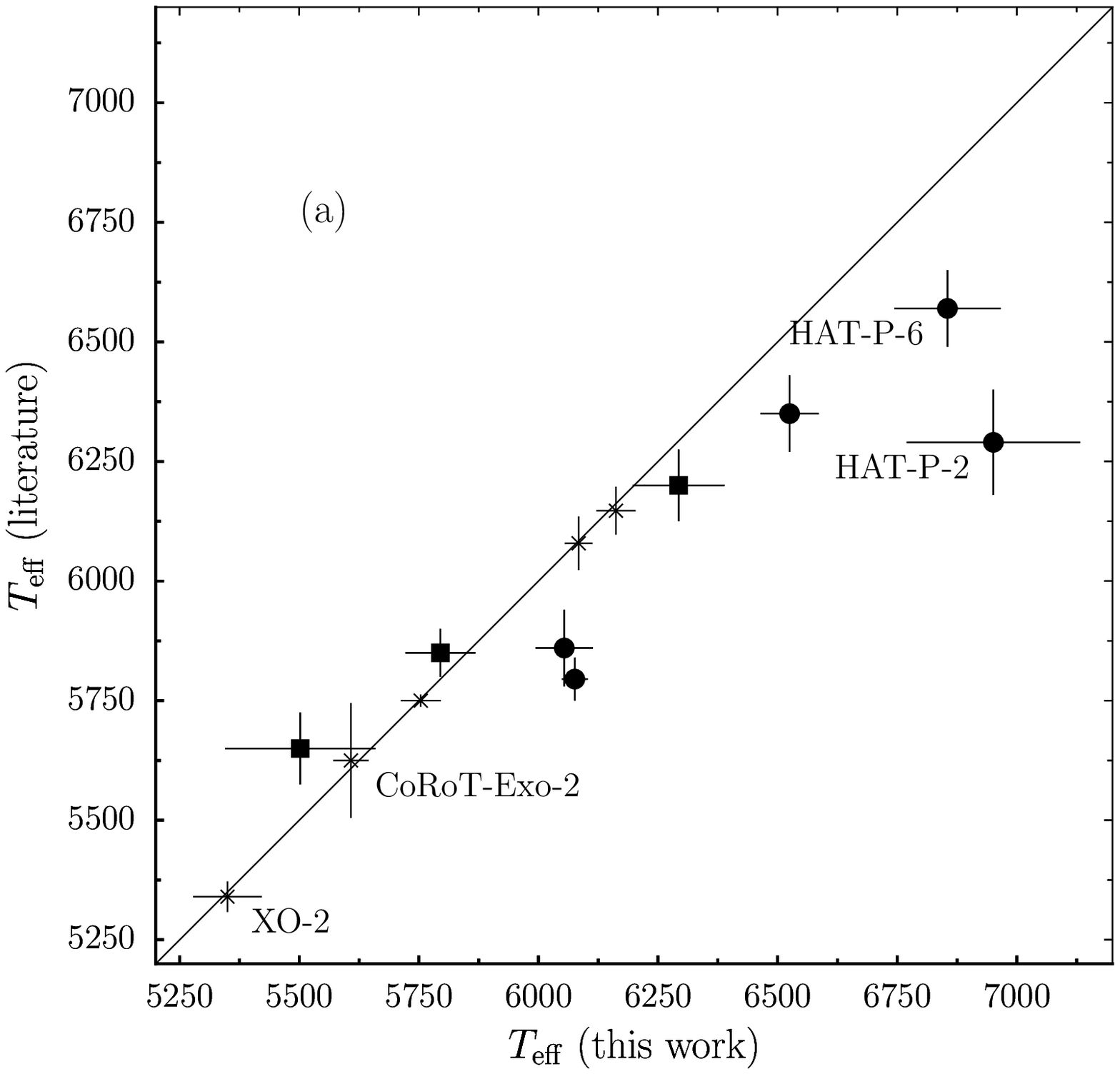}}
\subfigure{\includegraphics[width=88mm]{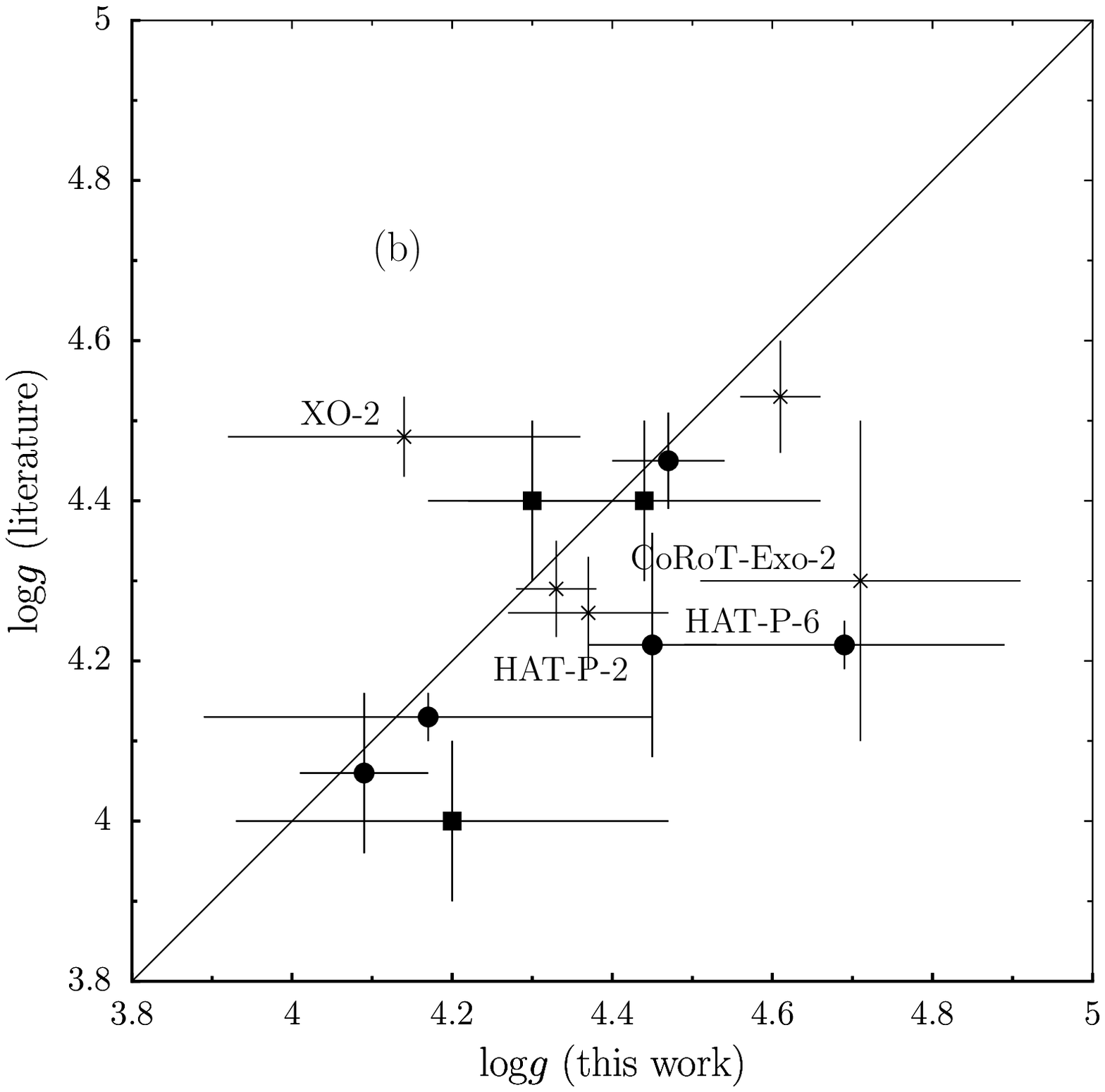}}\\
\subfigure{\includegraphics[width=88mm]{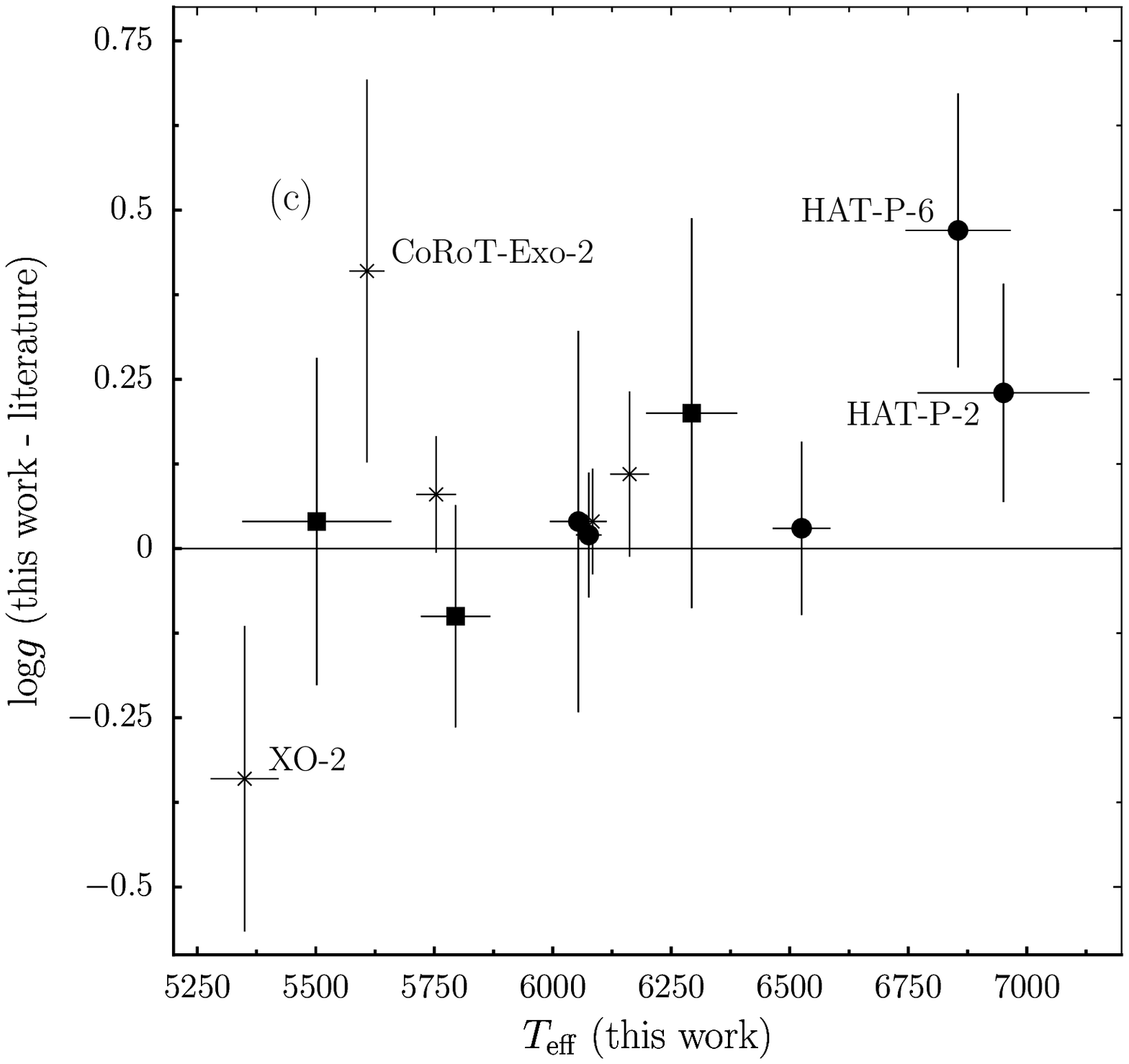}}
\subfigure{\includegraphics[width=88mm]{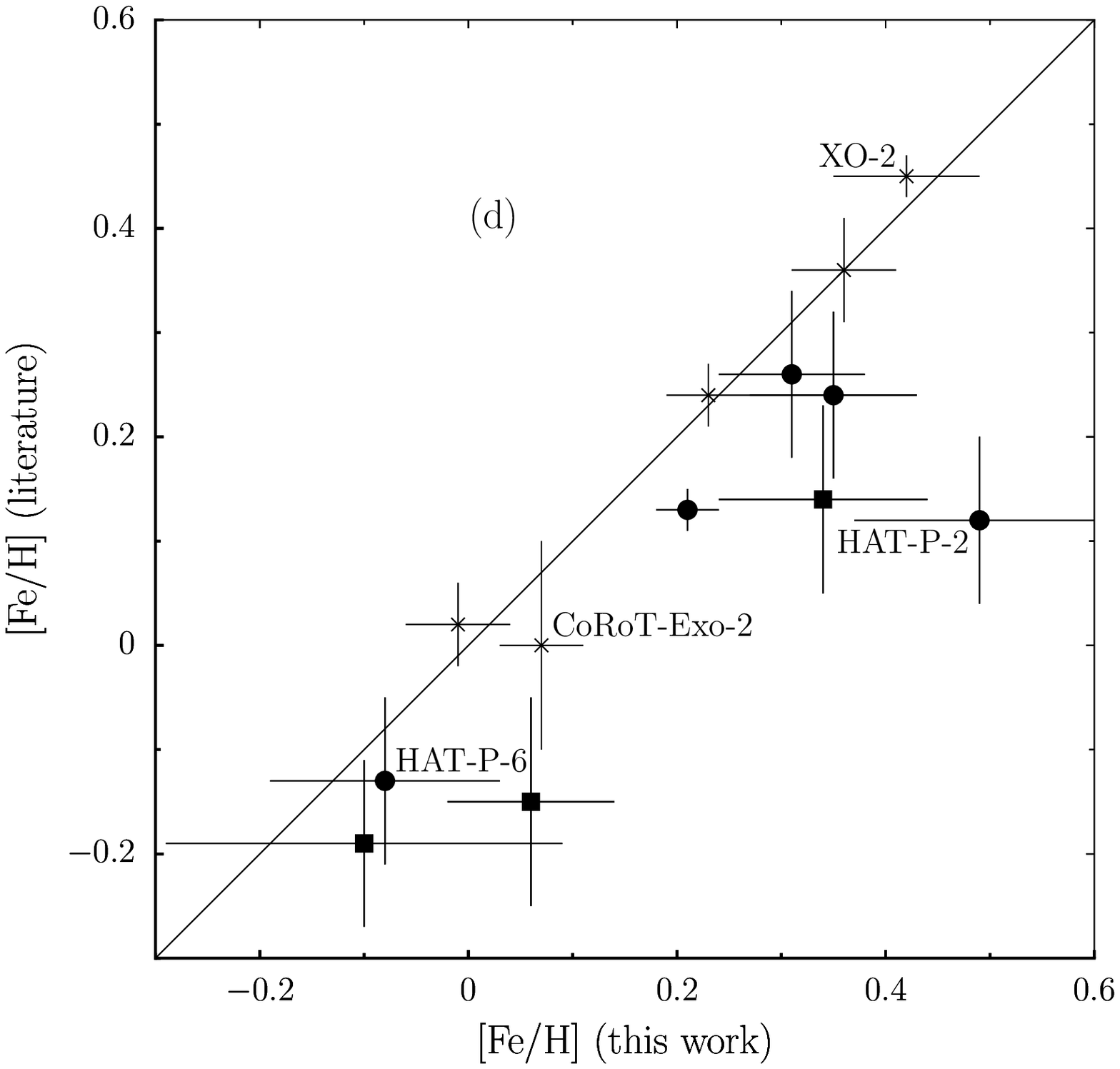}}
\caption{\label{fig:comp} {\bf Comparison with previous work.} -- Stellar parameters and iron abundance derived in the present work are compared with literature values. HAT objects are indicated by filled circles and TrES objects by squares. All other objects are depicted by crosses. Identity is marked by the solid line. Labelled data points are extensively discussed in the text. -- {\bf (a)} Effective temperatures obtained in the present work are compared with literature values. -- {\bf (b)} Surface gravities derived in the present work are compared with literature values. -- {\bf (c)} Residual surface gravities are displayed with respect to effective temperature. -- {\bf (d)} Iron abundance inferred in the present work are compared with literature values.}
\end{figure*}

In general, the new results are consistent with previous measurements as can be realized from a comparison of Tables~\ref{tab:par} and \ref{tab:others} (see Fig.~\ref{fig:comp} for graphical representations). Only the properties of HAT-P-2 appear discrepant to a very significant extent. While \citet{Bakos07b} derived an effective temperature of $6290\,$K for HAT-P-2 and an iron abundance of 0.12\,dex, the present work infers a temperature that is higher by 700\,K and an iron abundance of 0.49\,dex (see Fig.~\ref{fig:comp}a and d). When considering the formal error bars of HAT-P-1, both effective temperature and iron abundance appear to be in disagreement, while surface gravity agrees. In this particular case, the formal error bars almost certainly underestimate the true uncertainties. Although this implies that all stellar parameter measurements of HAT-P-1 could be in closer agreement, the offsets between given pairs of measurements are found to be systematic for all HAT objects.

We admit that there may be problems with Fe\,I at effective temperatures higher than $\approx6500$\,K, which may cause an overestimation of effective temperature. We note, however, that the previously derived effective temperatures of all the HAT host stars are systematically lower than the values of the present work, even at solar temperatures. At this point, one realizes that all HAT host stars were observed with the same Keck HIRES configuration and were analysed with the same spectral fitting technique. Taking into account possible systematics in the previous analyses as well as in the present work, we thus conclude that the true effective temperature of HAT-P-2 and HAT-P-6 will be somewhat higher than the previous measurements, but not as high as the values derived in the present work.

The iron abundance, which is the quantity primarily addressed by the present work, is also affected by systematics. The previously measured abundances of the HAT objects are systematically lower. In abundance analyses it is common for the origin of these discrepancies to be almost unidentifiable. In the present case, one may only conclude that it is not caused entirely by the use of different analysis packages. If this would be the case, similar deviations would also be expected for some of the abundance data represented by crosses in Fig.~\ref{fig:comp}. As for the HAT objects, some of these are also based on the approach of \citet{Valenti05} but perfectly agree with the present work.

The consideration of the TrES objects (Fig.~\ref{fig:comp}, squares) also shows that the situation might be more complicated, since \citet{Sozzetti07,Sozzetti09a} analysed the TrES stars essentially the same way as in the present work, but using Keck HIRES spectra with a higher spectral resolution than in previous studies of the HAT stars. Inspection of Figs.~\ref{fig:comp}a and b shows that the previously derived stellar parameters of the TrES objects are in only slightly closer agreement with the present work than the literature values of the HAT objects. As for the HAT objects, the abundance determinations of \citet{Sozzetti07} and \citet{Sozzetti09a} are systematically lower than those derived in the present work using the methods of \citet{Santos06b}. This is true not only for TrES-2, 3, 4 but also for TrES-1, which was studied by \citet{Sozzetti06} and \citet{Santos06b}.

In general, discrepancies in surface gravity hardly affect iron abundance as long as effective temperatures are in agreement. This is exemplified markedly by HAT-P-2, HAT-P-6, CoRoT-Exo-2, and XO-2 in Fig.~\ref{fig:comp}. On the one hand, the reader may note a substantial difference in surface gravity for the last three objects (Fig.~\ref{fig:comp}b). However, there are far fewer discrepancies and even close agreement between the different values of temperature and iron abundance. On the other hand, HAT-P-2, which exhibits the greatest variation in its values of abundance, also exhibits the greatest variation in the measurements of effective temperature. While the iron abundances are found to be relatively insensitive to variations in surface gravity, the planetary parameters are usually not found to be so insensitive when the surface gravity of the host stars is used to derive them. This particular problem is overcome by applying the mean stellar density from the transit light curves \citep[e.g.,][]{Sozzetti07,Holman07,Torres08}, thus avoiding the use of highly uncertain spectroscopic surface gravities.
This workaround is not possible in abundance analysis and therefore, the present work addresses spectroscopic surface gravities whenever possible. The determination of chemical abundance is closely tied to the adopted values of effective temperature and surface gravity. Stellar parameters and abundances must fulfil the requirements of ionization and excitation equilibria in a consistent way. The unverified adoption of external values of effective temperatures (e.g., from colour index calibrations) or surface gravities (e.g., from evolutionary models) may lead to spurious abundances \citep{Affer05,Luck05}. 

\section{The metallicity distribution of stars harbouring transiting planets}
The results of the present work are now combined with previous analyses of host stars of transiting planets completed using the methods of \citet{Santos04} (upper part of Table~\ref{tab:par}). 
\begin{figure*}
\subfigure{\includegraphics[width=88mm]{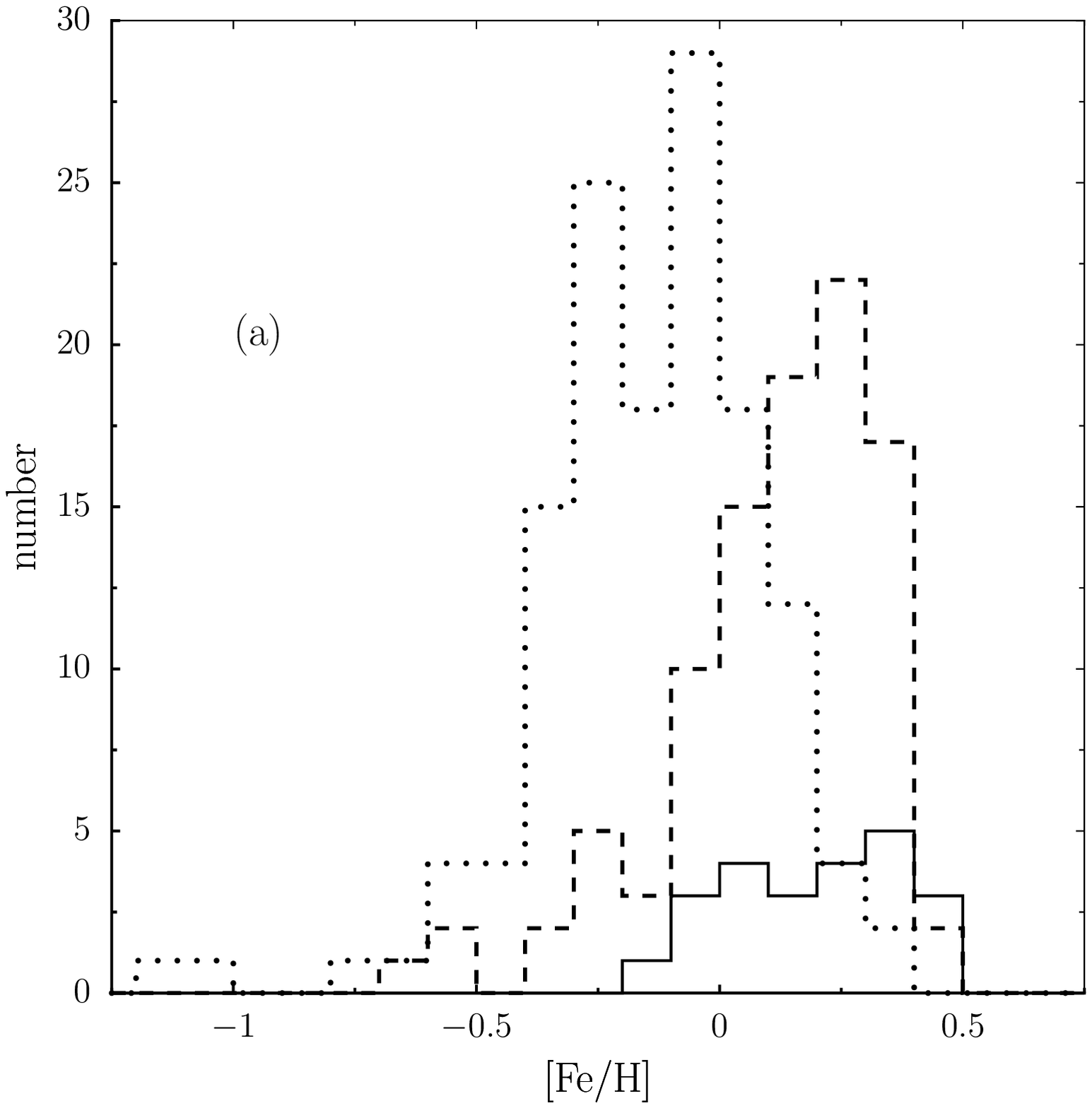}}
\subfigure{\includegraphics[width=88mm]{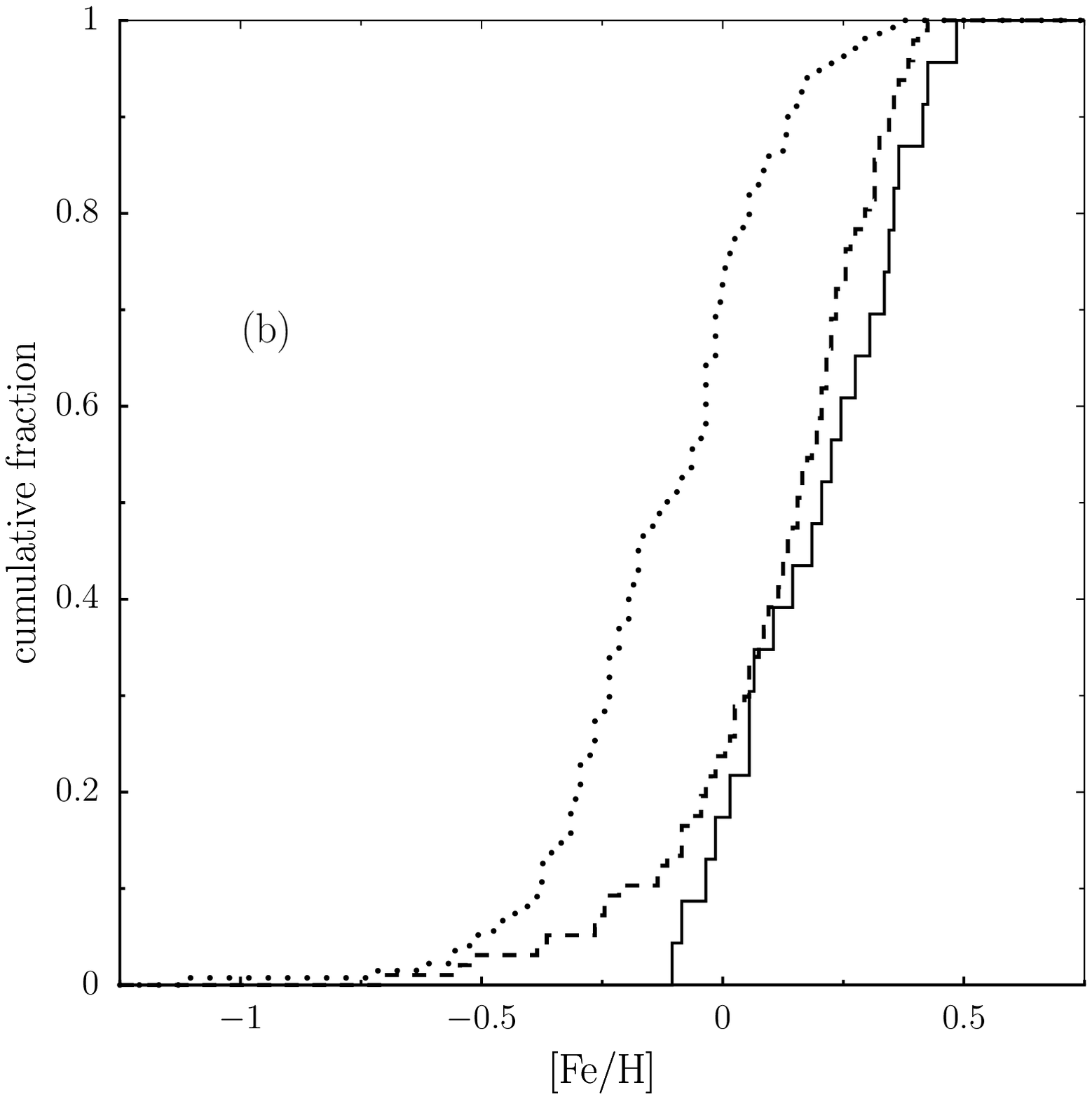}}
\caption{\label{fig:feh_dist} {\bf Comparison of the distribution of [Fe/H] of stars with and without planets.} -- {\bf (a)} The histogram displays the distribution of the stars with planets (\citealp{Santos04}; dashed), the stars with transiting planets (upper part of Table~\ref{tab:par}; solid), compared to a sample of stars without planets (\citealp{Santos04,Santos05}; dotted). -- {\bf (b)} The cumulative fraction is shown with the same line styles as in (a).}
\end{figure*}
Figure~\ref{fig:feh_dist} compares the distribution of iron abundance [Fe/H] for three different samples: the distribution for the stars with planets from \citet{Santos04}, for a comparison sample of stars without planets \citep{Santos04,Santos05}, and for the data of stars hosting transit planets. The distribution of the transit-planet host stars coincides well with the distribution of stars with planets from \citet{Santos04}. There could be a slight shift in the data for transit-planet host stars towards higher abundance.\footnote{In a discussion of the OGLE objects, \citet{Sozzetti04} and \citet{Santos06a,Santos06b} already noted the large distance of most of these objects, sampling Galactic regions of a possibly higher overall metallicity. The verification of this effect would require the analysis of many stars without planets in the same galactic regions.

}
However, the p-value of a K-S test is 0.27, which is consistent with the samples being drawn from the same distribution. One may note that the distribution of stars harbouring transiting planets does not show a tail towards low metallicities. This difference is not detected by the K-S test, which is less sensitive at the ends of the distributions. Still, the lack of a tail is not significant here because of the small number of measurements.

A comparison of the present work with the work of \citet{Santos04} can be completed accurately in a relative sense. All data shown in Fig.~\ref{fig:feh_dist} were derived using the same spectroscopic techniques. Therefore, any systematic offsets caused by different methods of analysis are not expected here.

\section{Summary and discussion}

We have derived the stellar parameters and iron abundances of 13 host stars of transiting planets. The analysed data consist of spectra obtained with the SARG, SOPHIE, and UVES spectrographs, and were analysed by assuming the iron ionization and excitation equilibria.

A comparison with previous determinations reveals some discrepancies between different values of effective temperatures and surface gravities, which are most striking in the cases of CoRoT-Exo-2, HAT-P-2, HAT-P-6, and XO-2. Systematics are also noted for the iron abundance measurements -- a finding which cannot easily be explained. Previous abundance measurements for the HAT and the TrES objects are systematically lower, while the data for other stars are in agreement with the present work. In general, abundances are found to be very robust with respect to possible uncertainties in surface gravity.

Although uniform data sets such as the literature data of the HAT and TrES objects as well as the present work principally allow one to address the origin of systematic discrepancies, this cannot be easily accomplished in the present case. The effects of instrumental configuration, reduction techniques, choice of atmospheric parameters, and line lists, cannot be determined separately with the available data sets.

The inferred iron abundance is affected most substantially in the case of HAT-P-2 (also known as HD\,147506), which also shows the largest temperature discrepancies. This is particularly interesting because the
$\sim 8\,M_\mathrm{Jup}$ planetary companion has been shown to be
anomalously small, with models fitting the inferred planetary radius
yielding suprisingly large enrichments in heavy elements (above
200\,$M_\oplus$; \citealp{Baraffe08}). The accretion of such a high mass in heavy
elements is difficult to explain with present formation models
\citep[see e.g.,][]{Ikoma06}. The improved effective temperature and metallicity
of the parent star found here opens up the possibility that updated stellar
parameters accounting for the new measurements yield a planetary radius
that is closer to expectations.

Extending the work of \citet{Santos06b}, an extensive compilation of homogeneously derived effective temperatures, surface gravities, and iron abundances of 24 stars harbouring transiting planets is now available.
Therefore, the distribution of the iron abundances can be compared very accurately in a relative sense, to the distribution of a large sample of planet host stars presented in \citet{Santos04}. The distributions are consistent with being drawn from the same population. 
The uniform data presented in this paper can be used for further statistical studies of the properties of transiting planet-host stars, and to refine a sample of uniform paramaters for the planets. This work will be the subject of a forthcoming paper.

The dataset was supplemented by spectroscopic parameters and in particular the iron abundance of 26 additional objects, thus comprising spectroscopic data of 50 host stars of transiting planets. Such a compilation is important to future planet studies involving abundance measurements of the host stars.

\begin{acknowledgements}
MAvE is supported by a scholarship (reference SFRH/BPD/26817/2006) granted by the Funda\c{c}\~ao para a Ci\^{e}ncia e a Tecnologia (FCT), Portugal. N.C.S. would like to thank the support from Funda\c{c}\~ao para a
Ci\^encia e a Tecnologia, Portugal, through programme Ci\^encia 2007 and
project grant reference PTDC/CTE-AST/66643/2006. S.G.S would like to acknowledge the support from FCT in the form of a grant SFRH/BD/17952/2004. We thank our anonymous referee for taking a critical point of view and giving fruitful suggestions. This research has made use of the SIMBAD database,
operated at CDS, Strasbourg, France, and NASA's Astrophysics Data System Bibliographic Services.\end{acknowledgements}

\bibliographystyle{aa}
\bibliography{./trans-plan_par}
\end{document}